\documentclass[a4paper,11pt]{article}
\usepackage{jheppub}
\usepackage[T1]{fontenc}
\usepackage{graphicx,epstopdf}
\usepackage{ulem}
\usepackage{epsfig}
\usepackage{color}
\usepackage{multirow}
\usepackage{epstopdf}
\usepackage{float} 
\usepackage{caption} 
\usepackage{lineno}
\usepackage{shuffle}
\usepackage{subcaption}

\graphicspath{{./dias/}}

\title{Double logarithmic contribution to Higgs pair production in the high-energy limit}

\author[a,b]{Zhenghong Hu,}
\author[a,b]{Tao Liu}

\affiliation[a]{Institute of High Energy Physics,  Chinese Academy of Sciences, Beijing 100049, China}
\affiliation[b]{University of Chinese Academy of Sciences, Beijing 100049, China}

\emailAdd{huzh@ihep.ac.cn}
\emailAdd{liutao86@ihep.ac.cn}

\abstract{We study the leading logarithmic QCD corrections to Higgs pair production in the high-energy limit, which originate from soft quark exchange in the Feynman diagram and are thus suppressed by the quark mass. 
It is found that triangle Feynman diagrams can be analyzed in the same way as single Higgs production mediated by bottom quarks~\cite{Liu:2017vkm,Liu:2018czl,Liu:2021chn}, while for diagrams of box type we cannot get an all-order result due to the many more double logarithmic structures they own.
In this paper we calculate the abelian corrections at three loops for the first time after obtaining the corresponding one- and two-loop leading logarithms, which are in total agreement with the analytical results in the literature.}

\begin{document}
\maketitle
\flushbottom

\section{Introduction}
After the discovery of Higgs boson in 2012, one of the primary goals of the Large Hadron Collider (LHC) is to determine Higgs self-couplings, which are key to understanding the electroweak symmetry breaking sector of the Standard Model (SM) and probing new physics. It is known that Higgs boson pair production via gluon fusion is the simplest process to obtain the information of triple-Higgs coupling and is expected to be observed at the high-luminosity LHC~\cite{LHCHiggsCrossSectionWorkingGroup:2016ypw}. 
On the theoretical side, there have been many efforts on the precise calculations of the differential and total cross sections for this process (see, e.g., refs.~\cite{Glover:1987nx,Plehn:1996wb, Borowka:2016ehy,Borowka:2016ypz,Baglio:2018lrj,Davies:2019dfy,Campbell:2024tqg,Heinrich:2017kxx,Jones:2017giv,Heinrich:2019bkc,Bagnaschi:2023rbx,DeFlorian:2018eng,deFlorian:2016uhr,Grigo:2015dia,Davies:2021kex,Davies:2023obx,Davies:2024znp, Grazzini:2018bsd,Chen:2019lzz,Chen:2019fhs,AH:2022elh, Bi:2023bnq,Heinrich:2024dnz}). 
The next-to-leading order (NLO) QCD corrections including full top quark mass dependence are calculated with the help of numerical methods. Other QCD corrections are usually obtained in the heavy-top limit, since the full analytical evaluation of high-loop Feynman integrals is technically quite challenging.
Besides the large top quark mass expansions, there are also studies at the NLO and NNLO on analytical approximations that are valid in certain kinematic regions, such as top pair threshold expansion~\cite{Grober:2017uho}, small Higgs mass transverse momentum expansion~\cite{Bonciani:2018omm} and small $m_t$ expansion in the high-energy limit~\cite{Davies:2018ood,Davies:2018qvx,Davies:2019dfy}.\footnote{Attempts on analytical evaluations of electroweak or light quark corrections to $gg\rightarrow hh$ have been made in refs.~\cite{Davies:2022ram, Davies:2023npk, Davies:2025wke,Bonetti:2025vfd, Hu:2025aeo}.} In this paper we focus on the high-energy limit, in which the masses $m_h, m_t$ and Mandelstam variables $s,t$ satisfy the relation $m_h^2 \ll m_t^2 \ll s, |t|$.

It is found that there are power-suppressed leading(double) logarithmic corrections in the scattering amplitudes involving massive particles in the limit of small mass or high energy~\cite{Kotsky:1997rq,Penin:2014msa,Melnikov:2016emg,Penin:2016wiw,Liu:2017axv}. In contrast to the traditional Sudakov logarithms~\cite{Sudakov:1954sw,Frenkel:1976bj,Mueller:1979ih,Collins:1980ih,
Sen:1981sd,Sterman:1986aj}, these kind of corrections are related to the effect of eikonal charge non-conservation with soft fermion exchange.
After factoring out the infrared(IR) divergent Sudakov logarithms, 
the same QCD color structure has been observed for the finite contribution in different physical processes, e.g., light quark mediated Higgs production $gg\rightarrow h$~\cite{Liu:2017vkm,Liu:2018czl,Liu:2021chn} and $gg\rightarrow hg$~\cite{Liu:2024tkc}{\footnote {Note that here the leading singularity subtraction operator differs from Catani's operator which is used in most cases. There are finite double logarithmic differences at one loop.}}. Subleading logarithms for the amplitude of single Higgs production have also been studied in refs.~\cite{Anastasiou:2020vkr,Liu:2020wbn,Liu:2022ajh,Hou:2025ovb}.  
Naively one may expect that the previous analysis could be directly extended to $gg\rightarrow hh$, while analytical results in ref.~\cite{Davies:2018qvx} and logarithmic analysis~\cite{Jaskiewicz:2024xkd} in the framework of soft-collinear effective theory show that it is not the case. 
Thus, the main aim of this work is to understand the origin of double logarithms of Higgs pair production up to NLO, which first appear at the level of next-to-leading power. Then we will turn to the corresponding abelian corrections at three loops.   

As discussed in refs.~\cite{Davies:2018ood,Davies:2018qvx}, a simple Taylor expansion can be performed for the smallest scale $m_h$ without bringing any IR divergences in the high-energy limit. 
The large logarithm considered in the amplitude is $\log\rho$ with the definition{\footnote{In the double logarithmic approximation $\log m_t^2/|t| \simeq \log m_t^2/s$.}} $\rho=m_t^2/s$. 
It is easy to find that the coefficients of the leading logarithms are rational functions of $s,t$ and $m_t^2$~\cite{Davies:2018qvx}, so $m_h$ can be safely set to be zero at the beginning of our analysis. 
For triangle Feynman diagrams involving the triple Higgs coupling, their amplitudes have the same kinematic structure as the amplitude of aforementioned single Higgs production after a proper normalization and the double logarithmic corrections can be obtained from ref.~\cite{Liu:2017vkm,Liu:2018czl,Liu:2021chn}{\footnote {The definition of $\rho$ should also be changed accordingly.}}. 
Here we only need to focus on the contributions from the Feynman diagrams of box type. Note that there are also Feynman diagrams containing two triangle fermion loops, which first show up at two loops and will be left for studies in the future.   

The remainder of the paper is organized as follows. In the next section we introduce the notations and then discuss the leading logarithms at one loop. In section 3 we show the momentum configurations which would generate the double logarithmic contributions at two loops. We find that most of these configurations cannot be obtained by dressing soft gluon on one-loop diagrams, which implies diagrams of higher loops would be more difficult to analysis. Thus, only the finite abelian corrections at three loops are calculated in this paper and they are discussed in section 4. Section 5 is our conclusion.

\section{Notations and one-loop corrections}
Due to Lorentz and gauge invariance, the matrix element ${\cal M}^{ab}$ of Higgs boson pair production process   
\begin{eqnarray}
 g(q_1)g(q_2)\rightarrow h(q_3)h(q_4),   
\end{eqnarray}
where $a,b$ are $SU(3)$ color indices of the external gluon, can be parameterized as
\begin{eqnarray}
  {\cal M}^{ab} = 
  \varepsilon_{1,\mu}\varepsilon_{2,\nu}
  {\cal M}^{\mu\nu,ab}
  =
  \varepsilon_{1,\mu}\varepsilon_{2,\nu}
  \delta^{ab} X_0 s 
  \left( F_1 A_1^{\mu\nu} + F_2 A_2^{\mu\nu} \right).
\end{eqnarray}
Here $A_1^{\mu\nu}$ and $A_2^{\mu\nu}$ are two independent Lorentz structures  
\begin{eqnarray}
  A_1^{\mu\nu} &=& g^{\mu\nu} - {\frac{1}{q_{12}}q_1^\nu q_2^\mu
  },\nonumber\\
  A_2^{\mu\nu} &=& g^{\mu\nu}
                   + \frac{1}{q_T^2 q_{12}}\left(
                   q_{33}    q_1^\nu q_2^\mu
                   - 2q_{23} q_1^\nu q_3^\mu
                   - 2q_{13} q_3^\nu q_2^\mu
                   + 2q_{12} q_3^\mu q_3^\nu \right) ,
\end{eqnarray}
with $q_{ij} = q_i \cdot q_j$ and $q_T^2 = (2q_{13}q_{23})/(q_{12}-q_{33})$. 
$X_0$ is defined to be $X_0 = \frac{G_F}{\sqrt{2}} \frac{\alpha_s(\mu)}{2\pi} T_F$, where $G_F$ is the Fermi's constant, $T_F=1/2$ and $\mu$ is the renormalization scale. Mandelstam variables $s,t$ are defined to be $s=(q_1 + q_2)^2$ and $t=(q_1 - q_3)^2$. 
All the physical information now is contained in the functions $F_{1,2}$, and it is convenient to decompose them into ``triangle'', ``box'' and ``double triangle'' form factors, which are represented by $F_{\rm tri}$, $F_{\rm box}$ and $F_{\rm dt}$ respectively. Subscript $\rm dt$ here denotes the contribution from Feynman diagrams containing two triangle sub-diagrams that are connected via one gluon propagator. Next, we define the expansion of the form factors as    
\begin{eqnarray}
  F &=& F^{(0)} + \frac{\alpha_s}{4\pi} F^{(1)}  + \left(\frac{\alpha_s}{4\pi}\right)^2 F^{(2)} + \cdots
  \,, \\
  F_1^{(0)} &=& \frac{3 m_h^2}{s-m_h^2} F^{(0)}_{\rm tri}+F^{(0)}_{\rm box1}\,,
  \nonumber\\
  F_2^{(0)} &=& F^{(0)}_{\rm box2}\,, \nonumber\\
  F_1^{(1)} &=& \frac{3 m_h^2}{s-m_h^2} F^{(1)}_{\rm tri}+F^{(1)}_{\rm box1}
                +F^{(1)}_{\rm dt1}\,, \nonumber\\
  F_2^{(1)} &=& F^{(1)}_{\rm box2}+F^{(1)}_{\rm dt2}\,, \nonumber \\
  F_1^{(2)} &=& \frac{3 m_h^2}{s-m_h^2} F^{(2)}_{\rm tri}+F^{(2)}_{\rm box1}
                +F^{(2)}_{\rm dt1}\,, \nonumber\\
  F_2^{(2)} &=& F^{(2)}_{\rm box2}+F^{(2)}_{\rm dt2}\,. 
\end{eqnarray}
These form factors could be expressed as linear combinations of scalar integrals with the help of known projectors~\cite{Borowka:2016ypz,Davies:2018ood}. 

As mentioned in the introduction there is no contribution to $F_2$ from triangle Feynman diagrams. In the considered high-energy limit with $m_t^2 \ll s, |t|$ and $m_h=0$, the power-suppressed leading logarithms of $F_{\rm tri}$ have the same property as single Higgs production. As for two-loop $F_{\rm dt}$, which are proportional to the product of two three-point integrals and are finite by themselves, it is also easy to get their double logarithmic corrections. 
While for $F_{\rm dt}$ at three loops, one have to consider logarithmic contributions of new types, since unlike $F_{\rm tri}$ the IR divergent part of $F_{\rm dt}$ and $F_{\rm box}$ could mix together.
So in order to fully understand the double logarithmic corrections of the whole process, we have to first scrutinize the logarithms in $F_{\rm box}$ which is the main aim of this paper.

\begin{figure}
 \centering
  \includegraphics[scale=0.5]{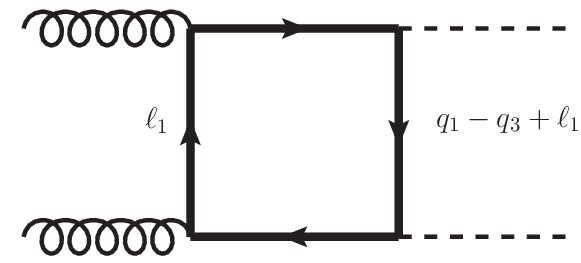}  
  \caption{One momentum configuration contributing to leading logarithms at one loop.}
\label{fig::1}
\end{figure}

Now let us turn to the one-loop contribution to $F_{\rm box}$. 
Sudakov's method is used for evaluating the leading logarithms as what we did before. Here the momentum configuration in Fig.~\ref{fig::1} is taken as an example to show the basic ideas. Soft quark momentum $l_1$ is parameterized by $l_1= q_1 u_1+q_2 v_1 + l_{1\perp}$ with the constraint $1>u_1v_1>m_t^2/s$ and two eikonal propagators are proportional to either $1/u_1$ or $1/v_1$. The denominator part of the soft and hard propagators reads  
 \begin{eqnarray}
    \frac{1}{\ell_1^2-m_t^2} &=& \frac{1}{2q_{12}u_1v_1 + \ell_{1\perp}^2 -m_t^2 },  \label{eq::soft}\\
    \frac{1}{(q_1-q_3+l_1)^2 -m_t^2} &=& \frac{1}{-2q_{13}} \left(1 + \frac{2(q_3-q_1)\cdot l_1 - l_1^2 + m_t^2 }{-2q_{13}} +  \cdots \right). \label{eq::hard}
\end{eqnarray}
Next we will discuss the mass suppression factor with a close look at the amplitude. 
Besides the $m_t^2$ factor provided by the Yukawa interactions, the numerator of soft quark propagator contribute either $m_t$ or $\ell_{1\perp}$ in the double logarithmic approximation, both of which will turn to an additional $m_t^2$ in the form factors. Due to the fact that $\ell_{1\perp}^2$ could be replaced by $m_t^2$ from Eq.~(\ref{eq::soft}), we have the following relation for $q_3 \cdot \ell_1$ 
\begin{equation}
     (q_3 \cdot \ell_{1})^2 =\frac{1}{2} q_{3\perp}^2 \ell_{1\perp}^2= -m_t^2\times\frac{q_{13}q_{23}}{q_{12}}. \label{eq::trans}
\end{equation}
Here $q_{3\perp}$ denotes the transverse component of $q_3$, which is parallel to $\ell_{1\perp}$.  
Note that besides the numerators in the amplitude, one can also get $q_3 \cdot \ell_{1}$ from expansion of the hard propagator as Eq.~(\ref{eq::hard}). 

The double logarithmic correction could be easily obtained after integrating over Sudakov variables. Obviously, each of the four quark propagators in this diagram can be soft and they all contribute the expected logarithms. E.g., if the hard momentum $q_1-q_3+\ell_1 $ becomes the soft one labeled by $\ell_1^\prime$, one should use the new decomposition $\ell_1^\prime= q_3 u_1 + q_4 v_1 + \ell_{1\perp}^\prime$ and Eqs.~(\ref{eq::soft},\ref{eq::hard},\ref{eq::trans}) should also be revised accordingly.     
So, there are four different configurations in every one-loop Feynman diagram of box type. Summing all these contribution gives 
\begin{equation}
 F^{(0)}_{\rm box1}=8\rho^2\ln^2\rho,~~~F^{(0)}_{\rm box2}=0,
\end{equation}
which are in agreement with ref.~\cite{Davies:2018qvx}. 

\section{Contribution at two loops}

\begin{figure}
\begin{center}
\begin{tabular}{ccc}
\includegraphics[scale=0.3]{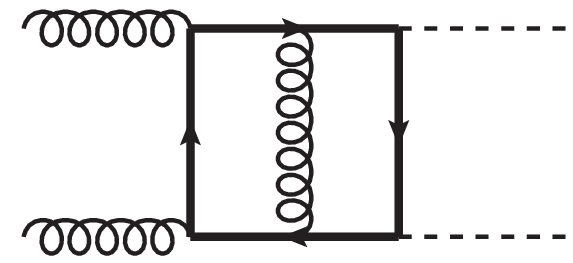} & \includegraphics[scale=0.3]{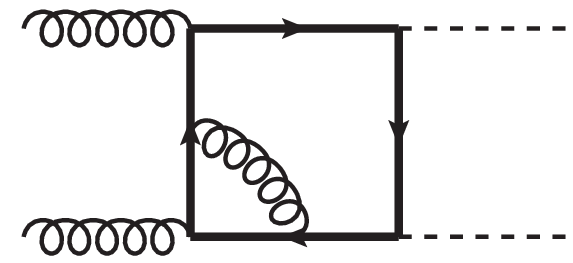} & \includegraphics[scale=0.3]{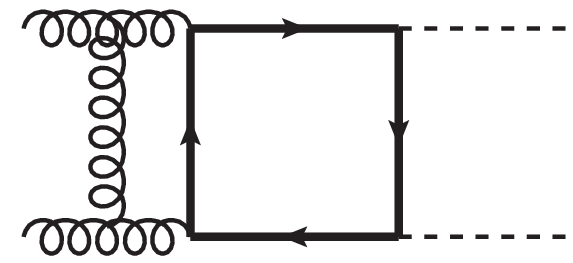} \\
(a) & (b) & (c)\\
\\
\includegraphics[scale=0.3]{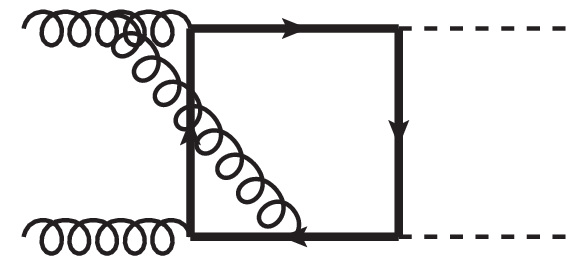} & \includegraphics[scale=0.3]{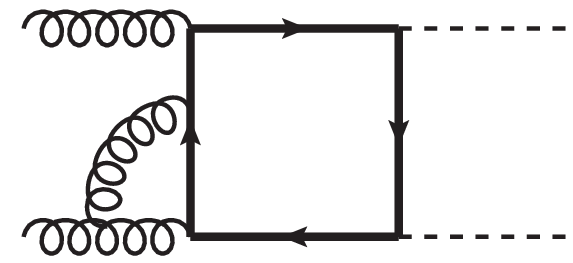} \\
(d) & (e) \\
\\
\includegraphics[scale=0.3]{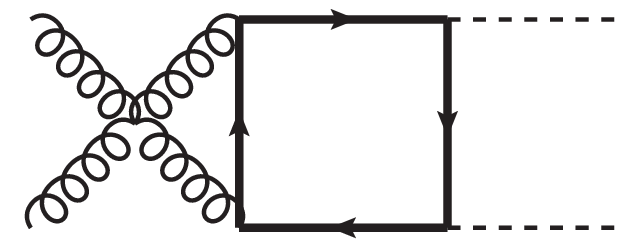}& \includegraphics[scale=0.3]{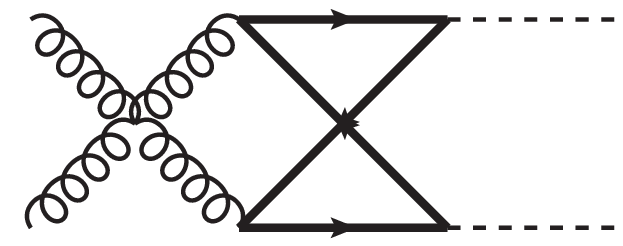}\\
(f) & (g) 
\end{tabular}
\end{center}
\caption{Typical two-loop Feynman diagrams contributing to double logarithmic correction.}
\label{fig::fig2}
\end{figure}

In contrast to the three-point amplitudes we have studied before~\cite{Liu:2017vkm,Liu:2018czl,Liu:2021chn}, dressing an additional soft gluon to the original one-loop diagrams only provides part of contribution at two loops. Later we will see that there are even extra eikonal factors $1/u_i$ or $1/v_i$ in certain new types of configurations, which need to be compensated by the soft momentum $\ell_i$ in the amplitudes before integration.
To get the correct double logarithmic contribution, one have to find all the suitable momentum configurations and then pick the relevant kinematic terms.          
In Fig.~\ref{fig::fig2} we list the typical Feynman diagrams contributing to the leading logarithms and the others could be easily obtained through permutations on the external legs of these samples.       

\begin{figure}
 \begin{center}
\begin{tabular}{ccc}
\includegraphics[scale=0.45]{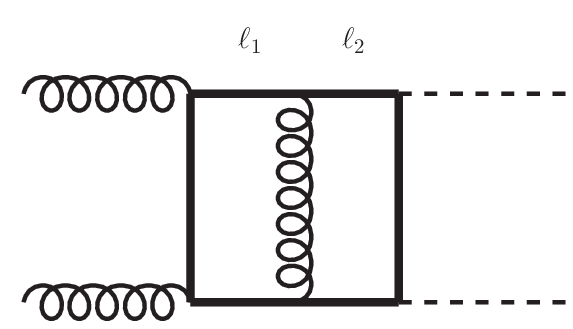} & \includegraphics[scale=0.45]{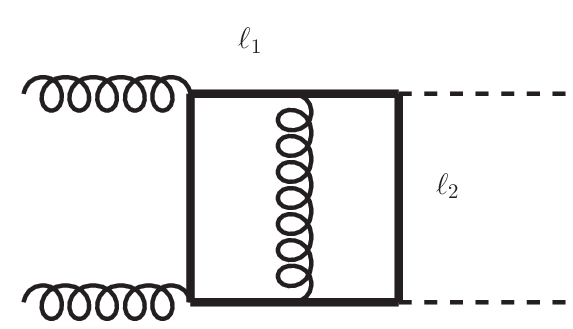} \\
($\rm a_1$) & ($\rm a_2$)\\
\\
\includegraphics[scale=0.45]{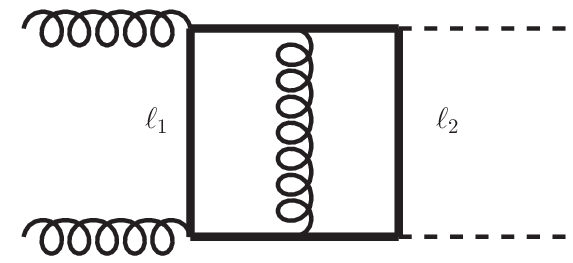} & \includegraphics[scale=0.45]{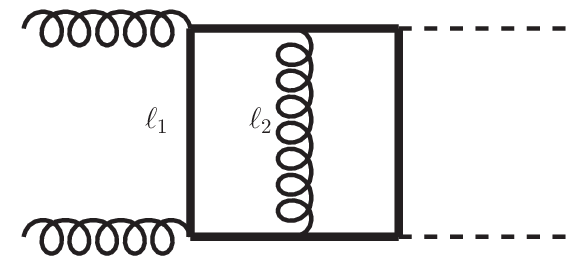}  \\
($\rm a_3$) & ($\rm a_4)$
\end{tabular}
\end{center}
\caption{Double logarithmic configurations of Fig.~\ref{fig::fig2}(a). Symmetric configurations are not shown.}
\label{fig::fig3}
\end{figure}

We start with Fig.~\ref{fig::fig2}(a) that owns nine double logarithmic momentum configurations, some of which are shown in Fig.~\ref{fig::fig3}. There are two soft quark lines and two hard lines in Fig.~\ref{fig::fig3}($\rm a_1$). If we parameterize $\ell_i= q_1 u_i + q_3 v_i + \ell_{i\perp}$, the middle gluon propagator will be proportional to 
\begin{eqnarray}
    \frac{1}{(\ell_1 -\ell_2)^2} = \frac{1}{-2q_{13}(u_1 v_2 + u_2 v_1) + \ell_{1\perp}^2  + \ell_{2\perp}^2 + \cdots}.
    \label{eq::douEi}
\end{eqnarray}
Having already two eikonal fermion lines which contain the eikonal factor $1/{(v_1 u_2)}$, one would immediately get two constraints for the above equation, which are $v_2 > v_1$ and $u_1 > v_2$. Since there are two soft fermion propagators, we don't need to expand the hard propagator to get the mass suppression factor when considering the logarithmic corrections of ${\cal O}(m_t^4)$. Note that the $q_1$ component of $\ell_1$ together with $q_3$ component of $\ell_2$ would provide $u_1 v_2$, which can be transformed into $\ell_{i\perp}^2$ during the expansion of Eq.~(\ref{eq::douEi}) and then into $m_t^2$ again as before. The next momentum configuration to be discussed is Fig.~\ref{fig::fig3}($\rm a_2$). The soft momenta lie between different external legs, so they need to be decomposed in different ways 
\begin{eqnarray}
    \ell_1= q_3 u_1 + q_1 v_1 + \ell_{1\perp}, \nonumber \\ 
    \ell_2= q_3 u_2 + q_4 v_2 + \ell_{2\perp}. \nonumber
\end{eqnarray}
Then the gluon propagator could be expanded as 
\begin{eqnarray}
    \frac{1}{(\ell_1 -\ell_2 +q_3)^2} =\frac{1}{2q_{13}v_1 + 2q_{34}v_2  -2q_1 \cdot \ell_2 v_1 + \cdots  }.
    \label{eq::a2}
\end{eqnarray}
Again, in order to get the double logarithmic contribution one obtains the condition $q_{13}v_1 > q_{34}v_2$, which can be simplified to $v_1 > v_2$ since the difference will only affect the next-to-leading logarithms in the high-energy limit. And it is also easy to check that there is no $m_t^2$ from the expansion of Eq.~(\ref{eq::a2}) or the hard propagator. The other two configurations in Fig.~\ref{fig::fig3} are simple and they will not be discussed.   

\begin{figure}
\begin{center}
\begin{tabular}{ccc}
\includegraphics[scale=0.45]{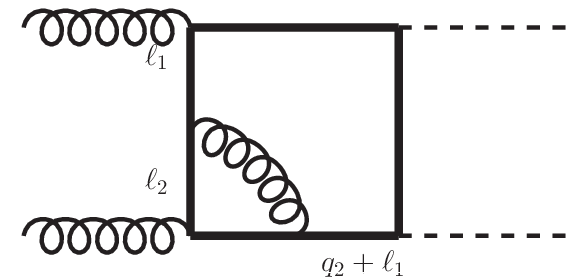} & \includegraphics[scale=0.45]{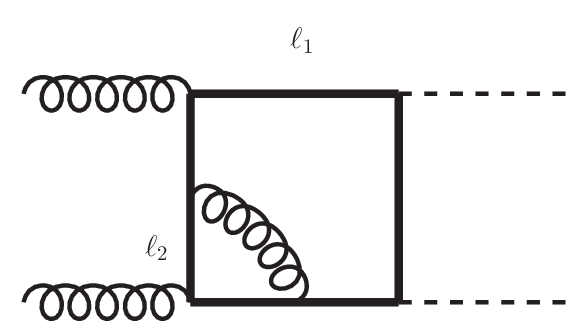} 
\\
($\rm b_1$) & ($\rm b_2$) 
\end{tabular}
\end{center}
\caption{Typical double logarithmic configurations of Fig.~\ref{fig::fig2}(b).}
\label{fig::fig4}
\end{figure}

Fig.~\ref{fig::fig2}(b) with two typical soft momentum distributions shown in Fig.~\ref{fig::fig4} seems easy to handle. Here we want to emphasize that there is an extra eikonal factor $1/(q_2\cdot\ell_1)$ in Fig.~\ref{fig::fig4}($\rm b_1$). This factor has to be canceled by the numerators in the Feynman integrals. There are also contributions from diagrams Fig.~\ref{fig::fig2}(f) and Fig.~\ref{fig::fig2}(g), which involve four gluon vertexes and are in essence three-point integrals. Fig.~\ref{fig::fig2}(f) has only one trivial region. Fig.~\ref{fig::fig2}(g) is found to have four non-vanishing logarithmic configurations during the calculation. 

Up to now all the contributions we have discussed are IR finite, actually the 
corrections could also be divergent when the momentum of gluon becomes soft, such as 
logarithmic regions of Fig.~\ref{fig::fig2}(c,e). These two Feynman diagrams have the same topology as Fig.~\ref{fig::fig2}(a,b), except different mass distributions for internal propagators. Thus, the momentum configurations we found earlier can be safely used for them. The IR divergent factor which could be factorized is exactly the same as the physical process of light quark mediated single Higgs production. As done in refs.~\cite{Liu:2017vkm,Liu:2018czl,Liu:2021chn} IR divergences are subtracted in the factorized form of simple integrals, then the finite correction can be compared with the results provided in ref.~\cite{Davies:2018qvx}.

\begin{figure}
\begin{center}
\includegraphics[scale=0.45]{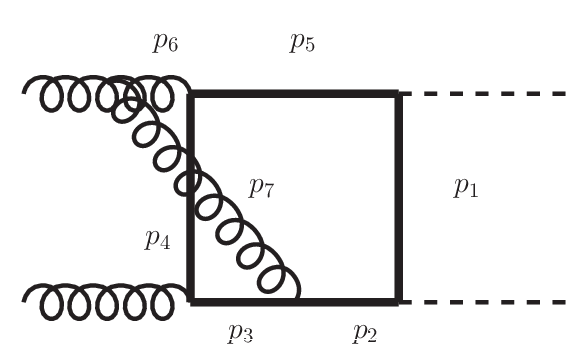}
\end{center}
\caption{To illustrate the double logarithmic configurations of Fig.~\ref{fig::fig2}(d), propagators of this diagram  are labeled as $p_i$ with index $i$ runs from 1 to 7.}
\label{fig::fig5}
\end{figure}

Having gained enough experience from the above analysis, we will turn to the most complicated non-planar Feynman diagram Fig.~\ref{fig::fig2}(d). To illustrate the 12 momentum configurations it owns,  we will label the propagators as $p_i$ with $i$ running from 1 to 7 in Fig.~\ref{fig::fig5}.
First, if $p_1$ is soft, there are four options for choosing the other soft momentum which are $p_{3,4,6,7}$. Needless to say, IR divergences come up when $p_6$ or $p_7$ becomes soft. 
The second type appears if the propagators $p_{3,6}$ or $p_{4,7}$ are soft. In both cases $p_1$ will be hard and could provides the transverse component of the soft momentum to obtain $m_t^2$ like Eq.~(\ref{eq::hard}). 
The combination of the soft momenta for the last six regions are $p_{45,46,56,23,27,37}$. Taking $p_{56}$ as an example, these two soft momenta could be decomposed with $q_3$ and $q_1$. With the same constraints on Sukakov variables as the configuration of Fig.~\ref{fig::fig3}($\rm a_1$), the propagator $p_4$ is proportional to $1/(u_1v_2)$ after simplifications. However, due to the presence of $p_3$,  one has to consider two kind of possibilities, i.e., $u_1>v_2$ or $v_2>u_1$. Different numerators in the amplitudes should be picked to cancel this extra eikonal factor in $p_3$\footnote{ %
This kind of problem has already been encountered in the analysis of  $gg\rightarrow hg$ \cite{Melnikov:2016emg}.}.  
Although not mentioned, one also has to be careful on the numerators 
which could be transformed to $m_t^2$, since the aforementioned transformation 
does not works for $\ell_{i\perp}^2$ if $\ell_i$ is the soft loop momentum of massless gluon. 

Before going to three loops, some comments are in order. To be honest, there are many more double logarithmic configurations than what we have expected at the beginning of this calculation. If only the abelian diagrams are taken into account, only Fig.~\ref{fig::fig3}($\rm a_4$) will be obtained via dressing a soft gluon on the one-loop configuration, which indicates that there should also be new kinds of contributions at three loops. On the other hand, in order to account for the mass factors in the amplitudes, sometimes we have to expand certain propagators to get the transverse component of the soft momentum $\ell_i$. In some configurations we need to do the expansion to compensate the extra eikonal factors and then to restore the double logarithmic structure in the Feynman integral. All these facts make the all-order analysis rather difficult, even only the abelian contribution is considered. Another point we want to emphasize is that both the form factors $F_{1,2}$ receive corrections from the box diagrams. Although the factorized IR divergent part is the traditional one which doesn't contain finite logarithmic terms as \cite{Liu:2024tkc}, it is still non-trivial to get the correct factorization due to the presence of the non-planar Feynman diagrams.   
Finally, the correct IR factorization and finite two-loop leading logarithmic contribution are obtained 
\begin{eqnarray}
 F^{(1)}_{\rm box1}&=&\left(\frac{-2C_F(3s^2+2st+2t^2)}{3t(s+t)}  +  \frac{C_A(3s^2+4st+4t^2)}{3t(s+t)}\right) \rho^2\ln^4\rho , \nonumber \\
 F^{(1)}_{\rm box2}&=&\left(\frac{-2C_F(s+2t)^2}{t(s+t)} + 
 \frac{C_A(3s^2+2st+2t^2)}{3t(s+t)}\right) \rho^2\ln^4\rho.
\end{eqnarray}
Here $C_F=(N_c^2-1)/(2N_c)$ and $C_A=N_c$ for non-abelian $SU(N_c)$ color group.  

\section{Abelian corrections at three loops}
\begin{figure}
\begin{center}
\begin{tabular}{ccc}
\includegraphics[scale=0.3]{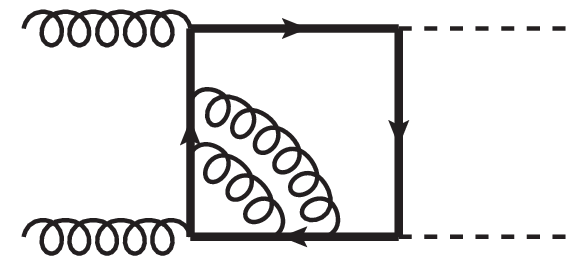} & \includegraphics[scale=0.3]{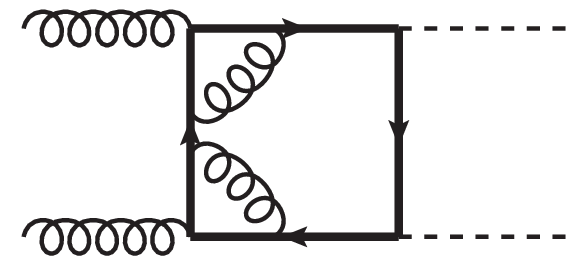} & \includegraphics[scale=0.3]{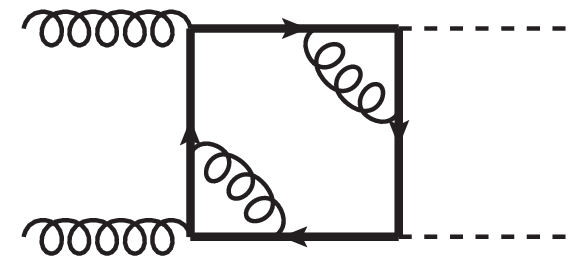} \\
(a) & (b) & (c)\\
\\
\includegraphics[scale=0.3]{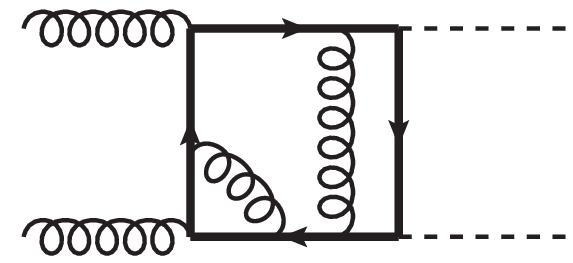} & \includegraphics[scale=0.3]{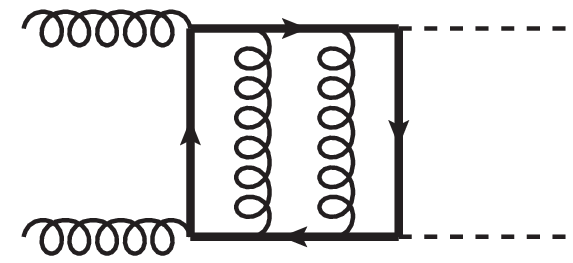} 
& \includegraphics[scale=0.3]{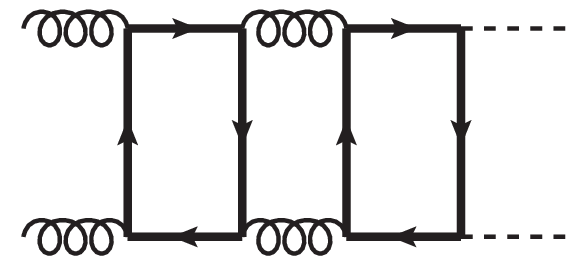}\\
(d) & (e) & (f) \\
\\
\\
\includegraphics[scale=0.3]{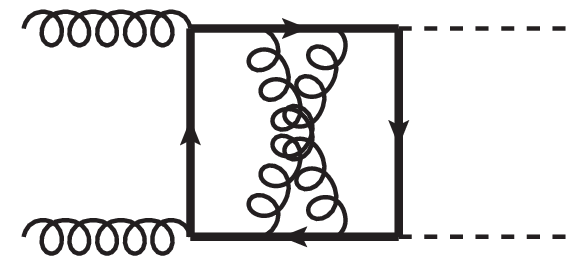}& \includegraphics[scale=0.3]{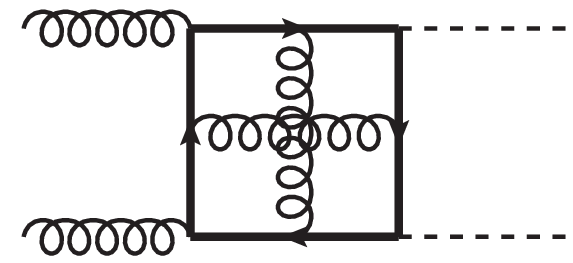} &
\includegraphics[scale=0.3]{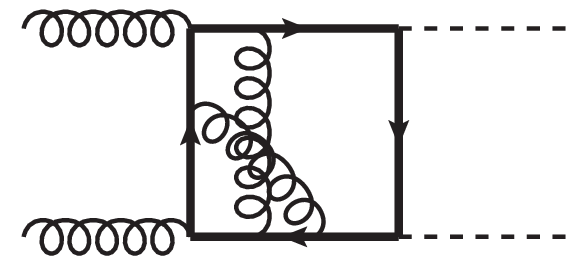} \\
(g) & (h) &(i) \\
\\
\includegraphics[scale=0.3]{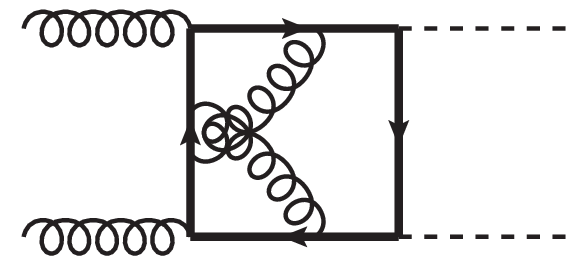} & \includegraphics[scale=0.3]{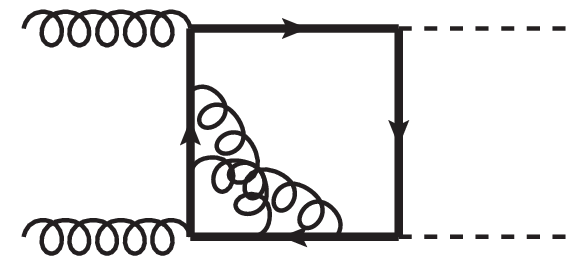} \\
(j) & (k) \\
\\
\includegraphics[scale=0.3]{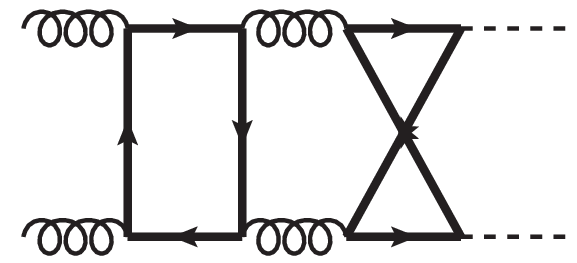}& \includegraphics[scale=0.3]{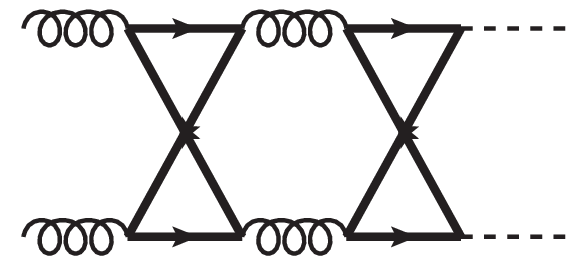}\\
(l) & (m) 
\end{tabular}
\end{center}
 \caption{Typical abelian Feynman diagrams contributing to double logarithmic corrections at three loops.}
 \label{fig::fig6}
\end{figure}

It is expected that evaluations at three loops would be much more complicated than before, since there are more Feynman diagrams and more double logarithmic momentum configurations in each diagram. Thus currently we only consider the abelian corrections which are IR finite. Following the same logic in the last section, 13 typical Feynman diagrams that contain the leading logarithmic regions are shown in Fig.~\ref{fig::fig6}. 
Although some diagrams or regions may not contribute the desired logarithms at ${\cal O}(m_t^4)$, they could be useful if higher expansion terms in $m_t$ or other four-point amplitudes in the high-energy limit are considered in the future. Also they might be useful for electroweak corrections to highly boosted Higgs pair production, the two-loop of which have been studied in refs.~\cite{Davies:2022ram,Davies:2023npk}. 

\begin{figure}
\begin{center}
\begin{tabular}{ccc}
 \includegraphics[scale=0.3]{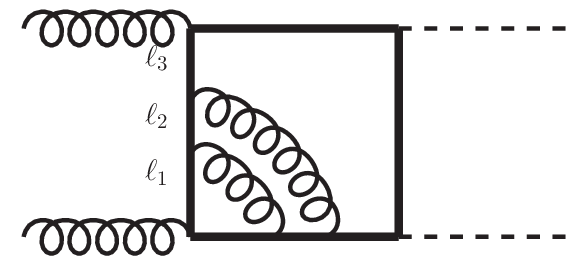} & \includegraphics[scale=0.3]{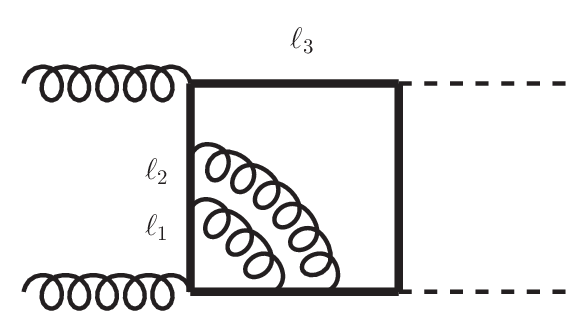} &
 \includegraphics[scale=0.3]{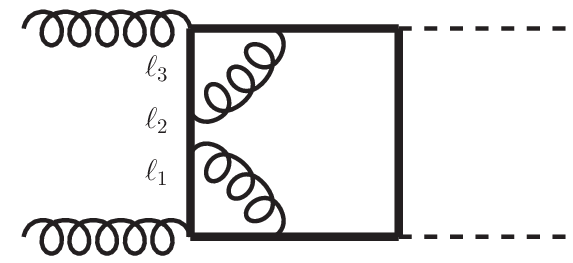} \\
 ($\rm a_1$) & ($\rm a_2$)  & ($\rm b_1$) \\
 \\
 \includegraphics[scale=0.3]{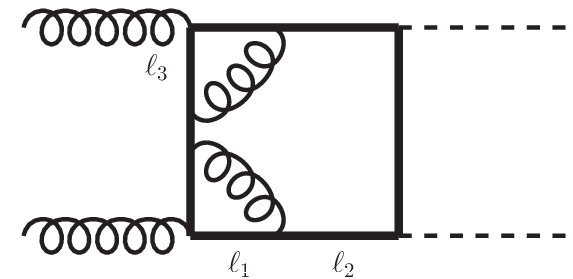} & \includegraphics[scale=0.3]{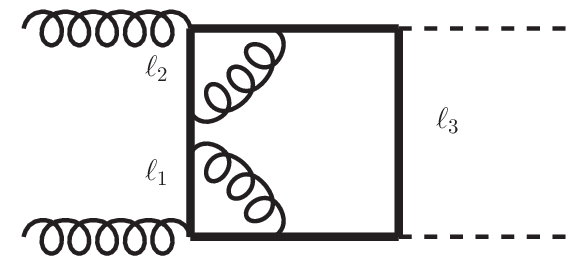} &
 \includegraphics[scale=0.3]{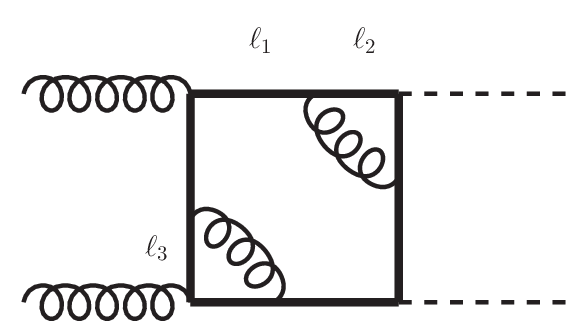} \\
 ($\rm b_2$) & ($\rm b_3$) & ($\rm c_1$) \\
 \\
 \includegraphics[scale=0.3]{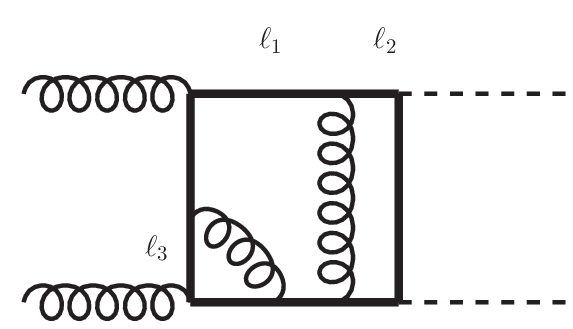} & \includegraphics[scale=0.3]{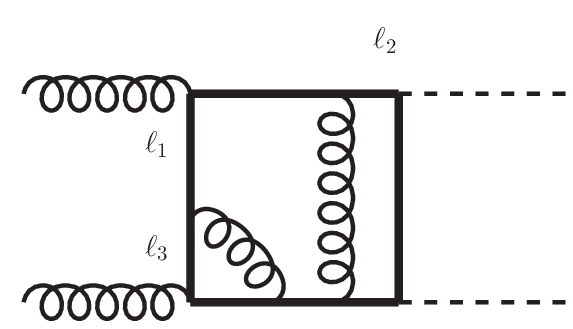} & \includegraphics[scale=0.3]{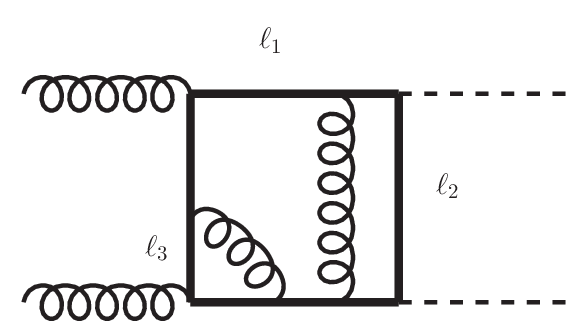} \\
 ($\rm d_1$) & ($\rm d_2$) & ($\rm d_3$) \\
 \\
 \includegraphics[scale=0.3]{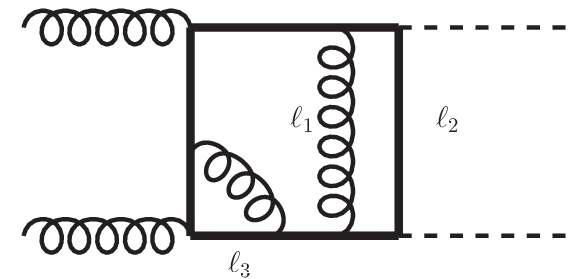} & \includegraphics[scale=0.3]{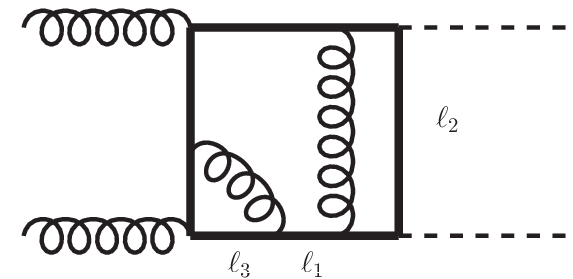} 
 & \includegraphics[scale=0.3]{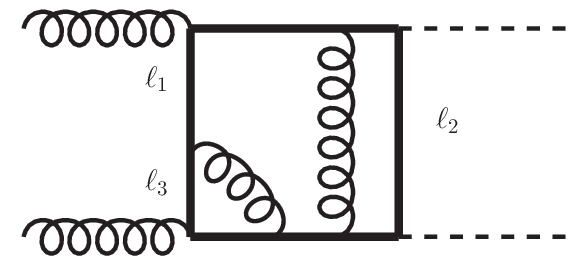}\\
 ($\rm d_4$) & ($\rm d_5$) & ($\rm d_6$) \\
 \\
 \includegraphics[scale=0.3]{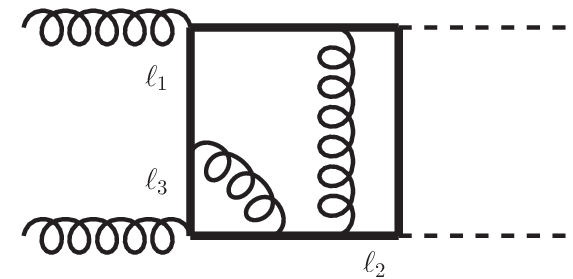}& \includegraphics[scale=0.3]{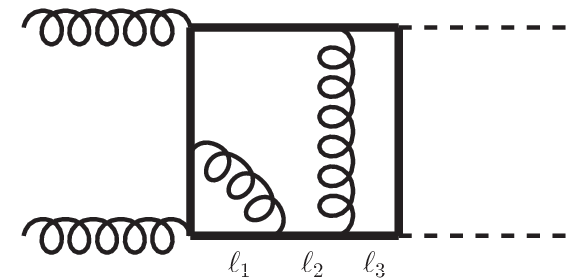} &
 \includegraphics[scale=0.3]{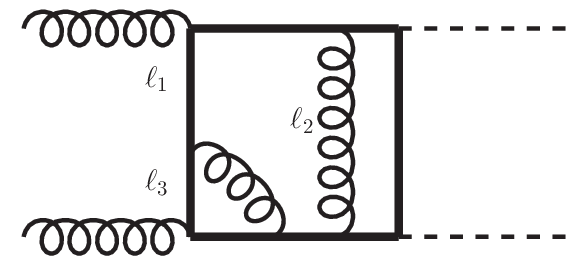}\\
 ($\rm d_7$) & ($\rm d_8$) & ($\rm d_9$) \\
 \\
 \includegraphics[scale=0.3]{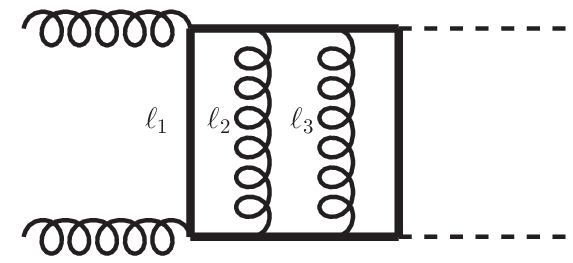} & \includegraphics[scale=0.3]{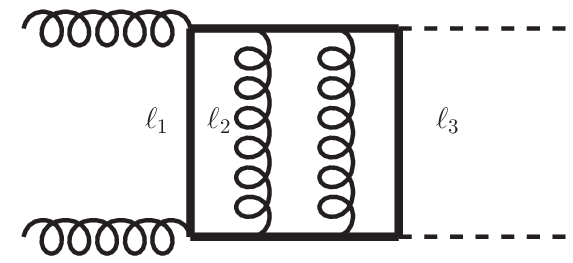} & 
 \includegraphics[scale=0.3]{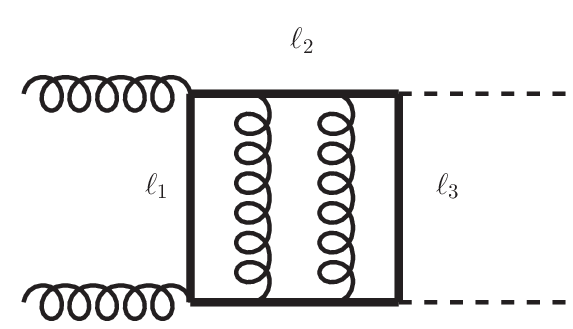}\\
 ($\rm e_1$) & ($\rm e_2$) & ($\rm e_3$) \\
 \\
 \includegraphics[scale=0.3]{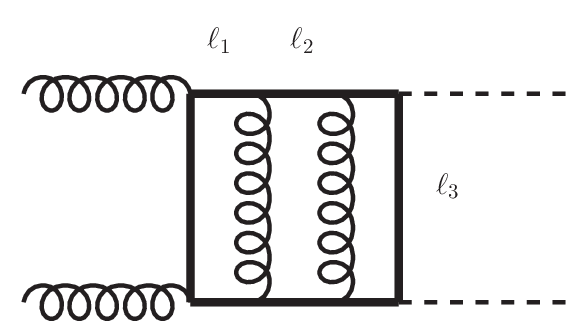} & \includegraphics[scale=0.3]{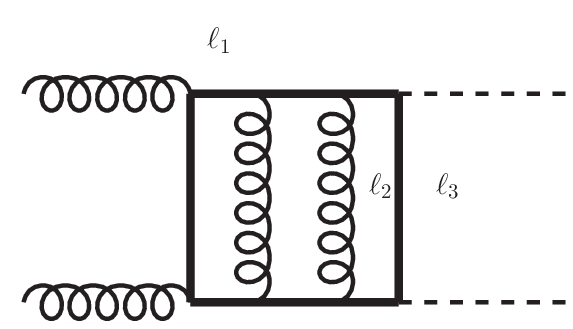} & 
 \includegraphics[scale=0.3]{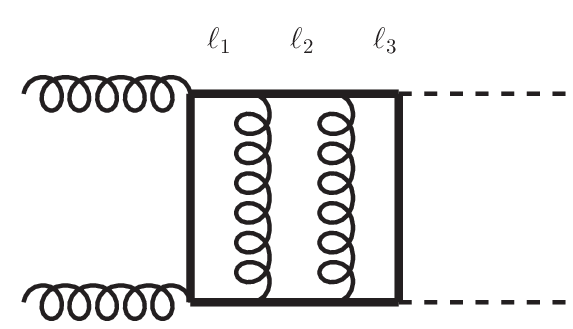} \\
 ($\rm e_4$) & ($\rm e_5$) & ($\rm e_6$) 
\end{tabular}
\end{center}
\caption{Typical leading momentum configurations of five planar Feynman diagrams in Fig.~\ref{fig::fig6}. $\ell_{1,2,3}$ denote the soft propagators.}
\label{fig::fig7}
\end{figure}

Comparing with the non-planar Feynman diagrams, planar ones are easy to analyze and will be discussed at the first step. 
Since Fig.~\ref{fig::fig6}(e) and Fig.~\ref{fig::fig6}(f) share the same topology with each other if top quark mass is neglected, 
it is enough to show all the typical momentum configurations for five typologies in Fig.~\ref{fig::fig7}. 
We need to keep in mind that here the unshown $\rm f_{1,...,6}$ have the same distributions as $\rm e_{1,...,6}$. 
It is easy to see that configurations with three soft quark propagators could be abandoned first, if there are no extra eikonal factors to be compensated.
Taken Feynman diagram Fig.~\ref{fig::fig6}(d) as an example, $\rm d_{1,2,3,5,6,7,8}$ of Fig.~\ref{fig::fig7} can not contribute the desired leading logarithms from this argument. 
As for $\rm d_9$, the contribution of which also vanishes from explicit calculations, it is interesting to see that it has the same structure as Fig.~4(c) of the ref.~\cite{Liu:2021chn} when the hard propagator of $\rm d_9$ shrinks to a point. 
Furthermore, we observe that similar non-vanishing correction from the non-planar configuration, which could be obtained by expanding the $h{\bar b}b$ vertex of Fig.~4(b) in ref.~\cite{Liu:2021chn} to one hard propagator. Although these two processes can not be compared directly, it may more or less explain why there is no contribution from $\rm d_9$. The last one $\rm d_4$ can be obtained by dressing a soft gluon to Fig.~\ref{fig::fig4}($\rm b_2$), and thus provides the expected logarithms.
Besides $\rm d_4$, non-vanishing configurations which are picked out for other Feynman diagrams are $\rm e_{1,2,5}$ and $\rm f_3$ in Fig.~\ref{fig::fig7}. $\rm e_{1,2,5}$ will not surprise us, while $\rm f_3$ is totally new and it first appears at three loops. After integrating out the transverse components $\ell_{i\perp}$, one will get the following relations of $u_i$ and $v_i$ for $\rm f_3$   
\begin{eqnarray}
 u_1 v_1 > \rho,~u_3 v_3 > \rho,~u_2 > u_1,~v_2 > u_3.  
\end{eqnarray}

\clearpage 
\begin{figure}
\begin{center}
\includegraphics[scale=0.5]{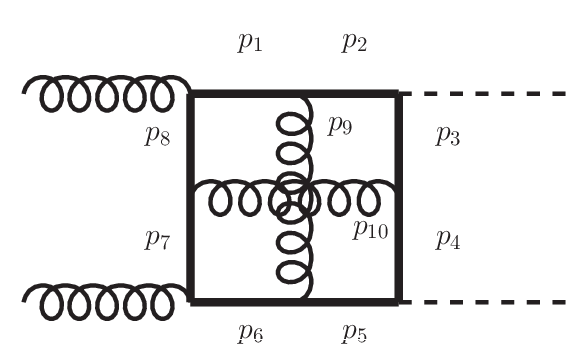}
\end{center}
\caption{Propagators of Fig.~\ref{fig::fig6}(h) are labeled by $p_i$ with $i$ runs from 1 to 10 to shown its double logarithmic configurations.}
\label{fig::fig8}
\end{figure}

\clearpage
As for the non-planar Feynman diagrams, we refrain from discussing them in detail, but take Fig.~\ref{fig::fig6}(h) as an example to show their complexities. Once again we label the propagators of this diagram to show its logarithmic configurations.   
In Fig.~\ref{fig::fig8} the configuration with propagators $p_{1,3,7}$ carrying the soft momenta could be discarded due to the same reason as Fig.~\ref{fig::fig6}($\rm d_1$). Although the scenario of soft propagators $p_{1,2,6}$ seems to work at the first glance, in real calculations it is found that the extra loop momentum from the soft quark momenta persists after proper transformations, and thus ruins the structure of leading logarithms.
The soft propagators of non-vanishing momentum regions and corresponding constraints on the Sudakov variables $u_i,v_i$ are given in Table.~\ref{tab::tab1}. For simplicity, we do not provide the detailed information on the decomposition of the soft loop momentum $l_i$, which should be easy to reconstruct.    
Here $p_{1,2,10}$ is the simplest region that one would expect from analysis at two loops. For $p_{1,2,7}$ we have the constraints $u_3> u_1$ and $u_3 > v_2$, 
since $\ell_3$ in the numerator have to canceled by the extra eikonal propagator containing one external momentum and three soft momenta. The other four configurations own two soft massive propagators and one soft massless propagator, so in order to compensate the extra eikonal factors the hard propagators have to be expanded. 
The numerators after expansion give the constraints on $u_i$ and $v_i$. 
Obviously, one will get different constraints with different overall coefficients if there are sufficient loop momentum in the numerator part of the Feynman integrals. 
Although we do not encounter this kind of problems here, it indeed exists during the calculation of Fig.~\ref{fig::fig6}(i). In summary, we found that there are 36 non-vanishing momentum configurations, which belong to six different classes, to calculate for this non-planar diagram.    


\begin{table}
\begin{center}
\begin{tabular}{ |c|c| } 
 \hline
Soft propagators of &\multirow{2}{*}{Constraints on $u_i$ and $v_i$ }   \\
logarithmic configurations  & ~ \\
\hline
$p_{1,2,7}$  &   $u_3>u_1>u_2,~~u_3>v_2>v_1, ~~v_3>v_1 $    \\
$p_{1,2,10}$ &   $u_1>u_2, ~~u_3>u_2, ~~v_3>v_1,~~v_2>v_1 $      \\
$p_{1,3,10}$ &   $u_2>u_3>u_1, ~~u_2>v_3>v_2, ~~v_1>v_2 $       \\
$p_{1,9,4}$  &   $u_1>u_2>u_3, ~~v_3>v_2>v_1, ~~u_1>v_2, ~~v_3>u_2 $   \\
$p_{1,9,6}$  &   $u_1>u_2>u_3, ~~v_3>v_2>v_1, ~~u_1>v_2, ~~v_3>u_2 $   \\
$p_{1,9,5}$  &   $u_1>u_2>u_3, ~~v_3>v_2>v_1, ~~v_3>u_1 $     \\
\hline
\end{tabular}
\end{center}
\caption{The soft propagators and constraints on Sudakov variables of double logarithmic configurations for Fig.~\ref{fig::fig8} are displayed in the table.}
\label{tab::tab1}
\end{table}

Similar to Fig.~\ref{fig::fig6}(h), there are many momentum configurations for other non-planar diagrams, most of which also start to contribute at three loops. Note that for each diagram the external legs have to be permuted to get all the contributions. These facts make the three-loop calculation more complicated than before. E.g., we obtain 24 diagrams belonging to the topology of Fig.~\ref{fig::fig6}(i) and each one owns more than 20 configurations to calculate.     
An interesting thing we found during the calculation is that the correction to Fig.\ref{fig::fig6}(l) and Fig.\ref{fig::fig6}(m) vanishes due to accidental cancellations between different configurations in each diagram, which can even be observed before integrating over Sudakov variables. Summing up all the leading logarithmic corrections from the non-planar and planar Feynman diagrams gives    
\begin{eqnarray}
 F^{(2)}_{\rm box1}&=&\left(\frac{-C_F^2(399s^2+628st+628t^2)}{720t(s+t)}  \right) \rho^2\ln^6\rho , \nonumber \\
 F^{(2)}_{\rm box2}&=&\left(\frac{-C_F^2(40s^2+43st+43t^2)}{180t(s+t)}\right) \rho^2\ln^6\rho, 
\end{eqnarray}
which can provide cross-checks for future analytical calculations. 

From the above analysis we observed new sources of leading logarithms at both two and three loops. Meanwhile, the number of new momentum configurations exceeds the number of known ones which could be obtained via dressing soft gluons between eikonal propagators in the known regions. It suggests that the double logarithmic structures could become more complicated at four or higher loops, even only the abelian corrections are considered. Thus novel methods may have to be developed for an all-order analysis, at least in the diagrammatic way we used. The results obtained in this paper can be considered as the first step toward this goal.

\section{Conclusion}
We calculated the leading logarithmic correction to Higgs boson pair production in the high-energy limit. The contribution from triangle Feynman diagrams could be obtained in the same way as the process of light quark mediated single Higgs production, while Feynman diagrams of box type own much more logarithmic momentum configurations. Furthermore, we found that there are new configurations at each loop in box diagrams and their numbers exceed the old ones. These facts, which never happen for three-point amplitudes, make the calculation rather complicated.  
The full corrections at one and two loops were obtained and they agree with the analytical results in the literature after subtracting the IR divergences. For three-loop contribution, we only calculated the finite abelian correction which could be cross-checked by future calculations. The non-abelian part at three loops which in principle should not be a big problem is left for future studies. From our analysis it is easy to see that a novel approach may have to be developed to get an all-order analysis of the leading logarithmic corrections of the box Feynman diagrams.   

\section*{Acknowledgments}
We would like to thank Alexander Penin for numerous discussions.  
The work of Z.H. and T.L. is supported in part by the National Natural Science Foundation of China (NNSFC) under grant No.~12375082 and No.~12135013, and by Institute of High Energy Physics (IHEP, CAS) under Grants No. Y9515570U1. 
The Feynman diagrams were drawn with the help of JaxoDraw~\cite{Binosi:2003yf}.

\end{document}